\documentclass[sigconf,nonacm]{acmart}
\AtBeginDocument{%
  }

\copyrightyear{2026}
\acmYear{2026}
\setcopyright{cc}
\setcctype{by-nc-nd}
\acmConference[SCA/HPCAsia 2026]{SCA/HPCAsia 2026: Supercomputing Asia and International Conference on High Performance Computing in Asia Pacific Region}{January 26--29, 2026}{Osaka, Japan}
\acmBooktitle{SCA/HPCAsia 2026: Supercomputing Asia and International Conference on High Performance Computing in Asia Pacific Region (SCA/HPCAsia 2026), January 26--29, 2026, Osaka, Japan}
\acmDOI{10.1145/3773656.3773683}
\acmISBN{979-8-4007-2067-3/2026/01}

\usepackage{listings}
\usepackage{xcolor}
\usepackage{placeins}
\lstdefinestyle{customc}{
    language=C,
    basicstyle=\ttfamily\footnotesize,  
    keywordstyle=\color{blue}\bfseries,
    commentstyle=\color{green!60!black},
    stringstyle=\color{red},
    numbers=left,
    numberstyle=\tiny\color{gray},
    stepnumber=1,
    numbersep=5pt,
    backgroundcolor=\color{gray!5},  
    showspaces=false,
    showstringspaces=false,
    showtabs=false,
    frame=tb,  
    rulecolor=\color{black},
    tabsize=2,
    captionpos=b,  
    breaklines=true,
    breakatwhitespace=false,
    escapeinside={(*@}{@*)},
    literate={∀}{{$\forall$}}1 {∈}{{$\in$}}1 {<-}{{$\leftarrow$}}1 {<++-}{{$\xleftarrow{+}$}}1 {<+-}{{$\overset{\raisebox{-0.5ex}{$\scriptscriptstyle+$}}{\leftarrow}$}}1,
    morekeywords={sp, xp, socket, param, const, function, parallel, let},
    aboveskip=3mm,
    belowskip=3mm
}

\newcommand{\todo}[1]{\textcolor{red}{(TODO: #1)}}
\newcommand{\algorithm}[0]{domain translation }

\newcommand{\algorithmpropernoun}[0]{Domain Translation}
\newcommand{\latency}[0]{\ensuremath{\tau }}
\newcommand{\gridpoints}[0]{\ensuremath{n }}
\newcommand{\ghostwidth}[0]{\ensuremath{g }}
\newcommand{\stencilrange}[0]{\ensuremath{p }}
\newcommand{\itertime}[0]{\ensuremath{t }}
\newcommand{\iterfreq}[0]{\ensuremath{f }}
\newcommand{\serialtime}[0]{\ensuremath{c }}
\newcommand{\dimension}[0]{\ensuremath{d }}
\newcommand{\fabriclen}[0]{\ensuremath{w }}

\newcommand{\latitude}[0]{\ensuremath{\phi}}
\newcommand{\longitude}[0]{\ensuremath{\lambda}}
\newcommand{\coriolis}[0]{\ensuremath{\ell}}

\newif\ifGB
\GBfalse

\begin{document}



\title{Beyond Exascale: Dataflow Domain Translation on~a~Cerebras~Cluster}

\thanks{Corresponding authors emails: michael@cerebras.net, srajama@sandia.gov}

\author{Tomas Oppelstrup}
\affiliation{
 \institution{Cerebras Systems}
 \city{Sunnyvale}
 \state{CA}
 \country{USA}
}

\author{Nicholas Giamblanco}
\affiliation{
 \institution{Cerebras Systems}
 \city{Vancouver}
 \state{BC}
 \country{Canada}
}

\author{Delyan Z. Kalchev}
\affiliation{
 \institution{Cerebras Systems}
 \city{Sunnyvale}
 \state{CA}
 \country{USA}
}

\author{Ilya Sharapov}
\affiliation{
 \institution{Cerebras Systems}
 \city{Sunnyvale}
 \state{CA}
 \country{USA}
}

\author{Mark Taylor}
\affiliation{
 \institution{Sandia National Laboratories}
 \city{Albuquerque}
 \state{NM}
 \country{USA}
}

\author{Dirk Van Essendelft}
\affiliation{
 \institution{National Energy Technology Laboratory}
 \city{Morgantown}
 \state{WV}
 \country{USA}
}

\author{Sivasankaran Rajamanickam}
\affiliation{
 \institution{Sandia National Laboratories}
 \city{Albuquerque}
 \state{NM}
 \country{USA}
}

\author{Michael James}
\affiliation{
 \institution{Cerebras Systems}
 \city{Sunnyvale}
 \state{CA}
 \country{USA}
}

\renewcommand{\shortauthors}{Oppelstrup et al.}


\begin{abstract}
Simulation of physical systems is essential across scientific and engineering domains.
Commonly used domain decomposition methods are unable to simultaneously deliver both high simulation rate and high utilization in network computing environments.
In particular, Exascale systems deliver only a small fraction their peak performance for these workloads.
This paper introduces the novel \algorithmpropernoun{} algorithm, designed to overcome these limitations.
On a cluster of 64 Cerebras CS-3 systems, we use this method to demonstrate unprecedented cluster performance across a range of metrics:
we show simulations running in excess of 1.6 million time steps per second;
we also demonstrate perfect weak scaling at 88\% of peak performance.
At this cluster scale, our implementation provides 112 PFLOP/s in a power-unconstrained environment, and 57 GFLOP/J in a power-limited environment.
We illustrate the method by applying the shallow-water equations to model a tsunami following an asteroid impact at 460m-resolution on a planetary scale.\enlargethispage{15pt}


\end{abstract}

\keywords{Cluster Computing, Time Integration, Stencil Computations, Finite Difference Methods, Shallow Water Equations}


\maketitle

© 2026 by the Authors. This is the authors' version of the work. The definitive Version of Record was published in SCA/HPCAsia 2026, 
http://dx.doi.org/10.1145/3773656.3773683.

\begin{figure*}
  \centering
  \includegraphics[width=1.0\textwidth]{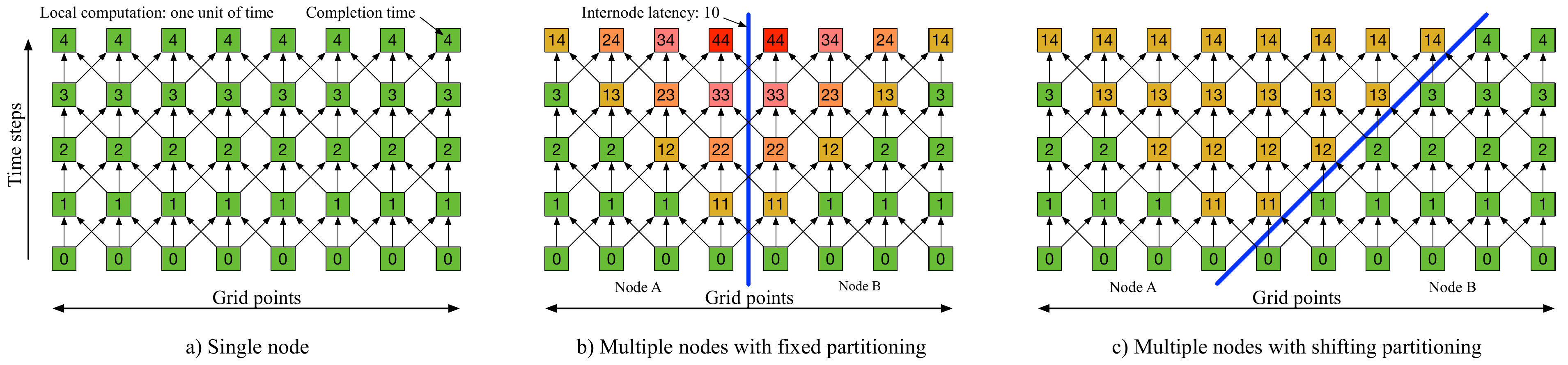}
  \caption{Illustrative example: a three point stencil in one dimension.
  (a) On single node there are no external dependencies and each time step takes one unit of time.
  (b) Static domain decomposition between nodes imposes an extra latency for communicating across node boundary (10 units in this example).  Grid points adjacent to the boundary experience additive latency delay at every time step. 
  (c) If the partition shifts by one unit at each time step, the cross-node latency is applied only once because the dependency across the domain boundary is unidirectional.  
  }
  \label{fig:motivation}
  \Description{Motivating example}
\end{figure*} 
\section{Introduction}

The ``principle of locality'' in physics asserts that objects are influenced only by their immediate surroundings.
This principle applies to both space and time.
It also forms the basis for Partial Differential Equations (PDEs) that relate rate of change at a point in space to local gradients.
PDEs describe various physical phenomena, including waves, potentials, diffusion, fluid motion, and electromagnetism. Numerical methods like finite differences, finite elements, and finite volumes, coupled with sufficient computing power, make it possible to study these equations at practical scales.


\enlargethispage{15pt}
These methods and ever-increasing computational power have driven the simulation of progressively larger systems using clusters of Von Neumann computers, culminating in the Exascale Computing Project that concluded in 2024 \cite{Exa_1, Exa_2}.  While the physical dimensions or resolution of the systems we could study grew by weak scaling, the temporal evolution rate has largely stalled due to the failure of strong scaling on Von Neumann architectures (i.e., the memory wall). Most typical Earth system models achieve less than 5\% of peak performance \cite{Govett_exa}.  Even with large peta- and exa-class machines, most large scale models achieve between 1.2 and 8 PFLOP/s, with the largest being 25.96 PFLOP/s \cite{Fu_camse,Ye_swnemo,fu2017solving,yang201610m,muller2019strong}.

Today, new computer architectures are addressing strong scaling. These massively parallel, spatial, and data-flow centric platforms are composed of a large number of Processing Elements (PEs) arranged in a grid and locally connected with a Network on Chip (NOC) router. Each PE resembles a small Von Neumann computer with a processor and local memory composed of SRAM. The Wafer Scale Engine (WSE) by Cerebras Systems\cite{HotChips}, the Dojo by Tesla\cite{Dojo_micro}, the Reconfigurable Dataflow Architecture by Sambanova\cite{Sambanova_micro} and the Tensor Streaming Processor by Groq\cite{Grok_Micro} are examples of this kind of architecture.  There is a natural resemblance between spatial architectures and the principle of locality in physics. Namely, when an object is represented in local PE memory, the influence of its surroundings becomes available with low latency and at speeds comparable to L1 cache access. The WSE is unique among these platforms as it is the only one that is manufactured at a full wafer scale with no chiplets or interposer giving it unique processing and power benefits. This was confirmed by previous studies that demonstrated that a single WSE node is highly efficient for evaluating PDEs in \cite{WSE_HPC_1, WSE_HPC_2}.

Here, we present the first distributed PDE solver on a cluster of WSEs. Further, we introduce a novel approach, based on the principle of locality, that maintains both physical and temporal locality across this spatial architecture so that networking latencies are completely hidden. We demonstrate the method by applying it to the solve two systems: (1) the heat equation (HE) with 5- and 9-point stencils, and (2) full-Earth tsunami simulations using Shallow Water Equations (SWE).

The SWE model horizontal fluid motion in liquids and gasses and multi-layer (stacked) SWE formulations are a key component of global atmosphere and ocean models used for both Earth system modeling and weather prediction. Moreover, the increasing frequency of asteroid detection in Earth's proximity \cite{JPL,EESA} inspired us to use SWE for simulating planetary-scale tsunami wave propagation caused by asteroid impacts in the ocean. Asteroids can dramatically impact life on Earth and even threaten its existence. Fast and accurate global simulations can enhance the understanding of such events and improve preparedness.

As we will demonstrate, this approach, in combination with the spatial architecture, is able to achieve perfect weak scaling at all tested processor workloads and maintain high efficiency down to just 256 elements per processor. Although this method is data-intensive, our three distinct codes can reach 57\%-88\% of computational peak utilization, even on a large cluster. Reaching 88\% of system peak is unprecedented for stencil computations. 

The rest of the paper is organized as follows. In the next section we give a brief overview of Domain Decomposition methods. Section 3 introduces Domain Translation, the new latency-hiding algorithm, discusses its properties, and compares it to traditional methods.  Section 4 describes the implementation of the method on a cluster of Cerebras WSE systems.  Section 5 characterizes the performance of the method applied to the heat and shallow water  equations.  Finally, the conclusion summarizes the results and describes how they advance the state of the art. 



\section{Distributed computations with domain decomposition}

Cluster implementations of geometry-based simulations typically use the domain decomposition method with different parts of the discretized mesh mapped to different nodes.  These simulations, including stencil computations, tend to have low operational intensity \cite{10.1145/1498765.1498785}, which makes it challenging to minimize the effects of communication in distributed implementations.  Special cases, such as Krylov solvers, can benefit from communication avoiding techniques \cite{DBLP:conf/ipps/DemmelHMY08}.  These specialized techniques use redundant computation to minimize data exchange between communicating nodes~\cite{DBLP:phd/basesearch/Hoemmen10}. 

\begin{table}
  \centering
    \small
    \begin{tabular}{ccl}
        \hline
        Symbol & Unit & Description \\
        \hline
        \latency      & s     & Latency - Time of flight on network link \\
        \gridpoints   & pts   & Grid points - Linear span of grid points \\
        \ghostwidth   & pts   & Ghost width - Thickness of overlap region \\
        \stencilrange & pts   & Stencil reach - Manhattan stencil radius \\ 
        \itertime     & s     & Iteration time - Computing full time step \\
        \iterfreq     & Hz    & Iteration frequency - Rate of time stepping \\
        \serialtime   & s     & Serial time - Computing one grid point \\
        \dimension    &       & Dimension - Dimensionality of domain \\
        \fabriclen    & cores & Fabric Size - Linear span of core array \\
        \hline
        \latitude     &       & Latitude \\
        \longitude    &       & Longitude \\
        \coriolis     &       & Coriolis force \\
        \hline
    \end{tabular}
    \caption{Notation and variables}
    \label{table:symbols}
\vspace{-15pt}
\end{table}

A stencil computation's asymptotic iteration rate can not exceed the rate of the slowest grid point.
For a stencil with range \stencilrange, grid points exchange data within a \stencilrange-neighborhood.
In a direct fixed mapping for domain decomposition methods \cite{domain_decomposition}, this exchange causes grid points adjacent to a subdomain boundary to incur network latency on every step.
Thus, rate-limiting the entire simulation based on the slowest network link with latency {\latency} to $1/\latency$~(Figure~\ref{fig:motivation}b).

For distributed stencil computations, a common approach is to replicate a layer of \textit{ghost} points \cite{DBLP:conf/sc/DingH01} to adjacent nodes \cite{10.1145/1953611.1953615}.  Each node holds replicated ghost values that belong to its neighbors' subdomains. 
These ghost
values participate in the computation, but after each iteration the ghost values at the outermost extent have undefined values because
they do not have access to information from the neighboring node. For the ghost method to reach a target iteration rate {\iterfreq } in the presence of network latency, the ghost region's thickness {\ghostwidth } must be proportional to $\latency\iterfreq$.
When $\gridpoints$ is the largest feasible domain radius for a node to achieve \iterfreq, the volume-fraction of the ghost region for a \dimension-dimensional domain is $1-\left(\frac{\gridpoints-2\ghostwidth}{ \gridpoints}\right)^\dimension$. This implies utilization decreases as rapidly as a degree-\dimension{} polynomial in $\latency\iterfreq$. 




While the overlap method amortizes the internode latency over multiple elements, it always sacrifices efficiency as it relies on redundant re-computations. It also forces a trade-off: increasing the rate of time step processing requires increasing the size of the ghost region, reducing both computational and power efficiency. 

In contrast, this work demonstrates PDE solvers running on a large cluster in a compute-bound regime fully independent of inter-node network latency. 
The achievement is based on a novel domain translation algorithm and its implementation on the WSE. Our implementation allows strong-scaling up to  1.6 million timesteps per second across a cluster with $10 \mu s$ interconnect latency. 

\section{Domain Translation algorithm}
We introduce \textit{\algorithmpropernoun}, a parallel algorithm for computing a stencil code efficiently over high-latency network links.
We will develop the algorithm presentation in a one-dimensional setting (Figure~\ref{fig:motivation}a).
The same principles hold in higher dimensions.
The algorithm requires nodes be connected in a ring (torus).

Application of a stencil operator uses information from grid points within a distance $\pm\stencilrange$.
Normally, this means that network links connecting subdomains carry bi-directional network traffic originating from $\stencilrange$ points on each side of the link.

The \algorithm algorithm translates the mapping from grid-points to processors by \stencilrange{} grid points on each iteration.
From the perspective of a network's links, translation composed with the stencil operator exchange causes unidirectional traffic flow.
Grid data now travels $2\stencilrange$ in the direction of translation and never travels against this direction.

Adjacent nodes have distinct \textit{upstream} and \textit{downstream} relationships. 
This unidirectional flow avoids additive buildup of latency contributions (Figure~\ref{fig:motivation}c).  A given grid point only experiences network latency after it has crossed a node's entire subdomain.
In this way, the impact of network latency is amortized over the width of the subdomain.
In contrast, a fixed partitioning domain decomposition of a fixed subset of gridpoints would experience latency cost every iteration (Figure~\ref{fig:motivation}b). 

As we will show, network latency causes no utilization loss when the size of a subdomain exceeds a critical threshold. Only link bandwidth and computational performance limit the iteration rate.
We note the algorithm requires additional network bandwidth both because it supports higher iteration rates and because local grid-point state that does not participate in stencil state exchange transits network links to stay with its grid point as the domain translates.



\begin{figure*}[t]
  \centering
  \includegraphics[width=0.9\linewidth]{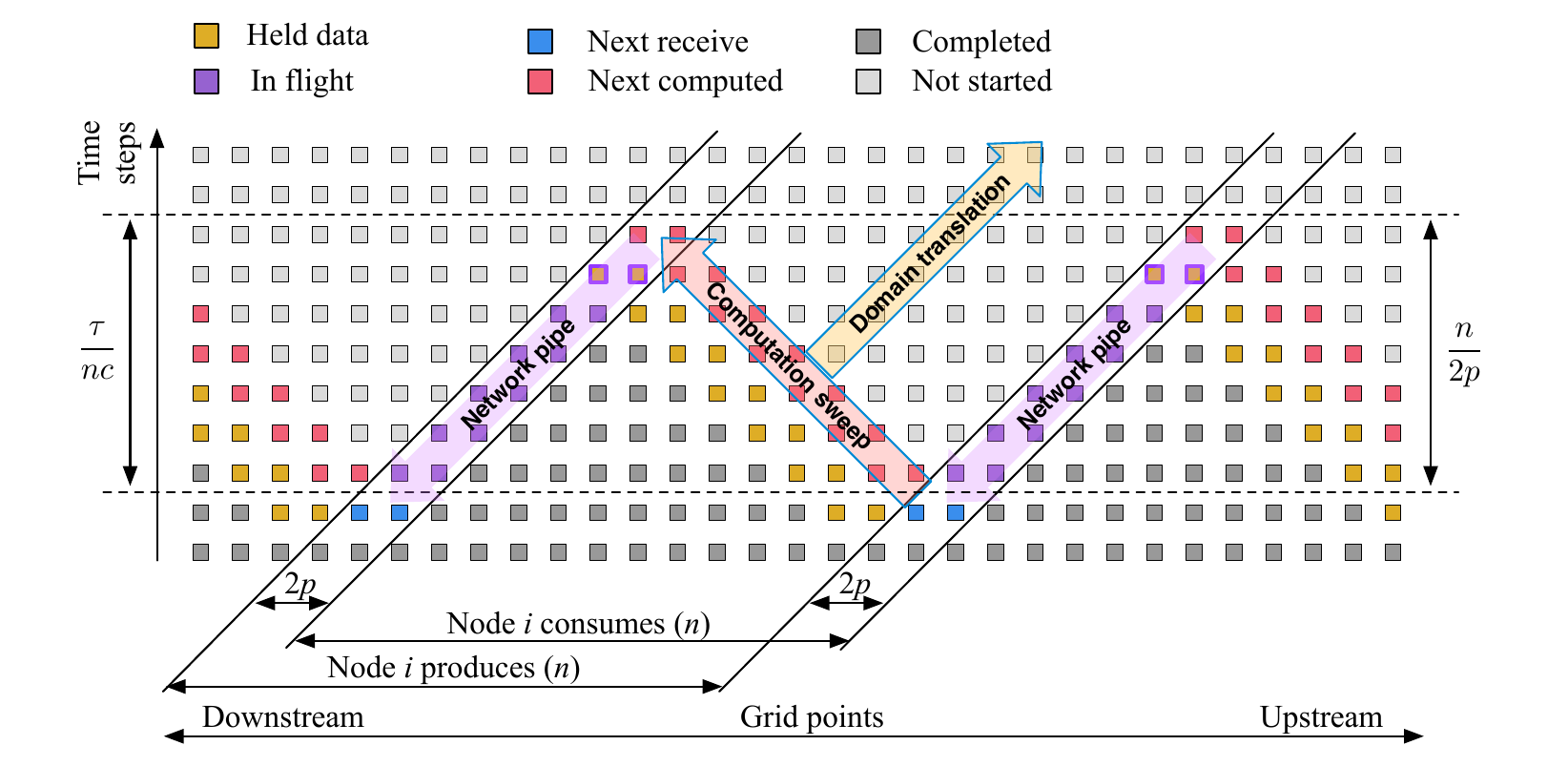}
  \caption{Domain translation method. Diagram shows grid points (x-axis) by timestep (y-axis). The diagonal lines indicate high-latency subdomain boundaries. The diagram depicts the algorithm's steady-state duty cycle. When the node receives a package (blue) it initiates a computation sweep that uses stored values (yellow) to produce the new grid point state (red). At the end of the sweep it submits its last computed values into the downstream network pipeline. The time used to complete the computation sweep is the same time it takes a network package to advance one ``step'' through the network pipeline. The end-to-end network latency coincides with $\gridpoints/2\stencilrange$ steps. The processor's wall-clock time proceeds in the direction of the ``domain translation'' arrow. Note that in the processor's time frame, it proceeds through a full network pipeline's worth of computation sweeps prior to receiving data that had just entered the upstream network pipeline.
  }
  \Description{Domain translation algorithm.}
  \label{fig:marching}
\end{figure*}

\begin{table*}
    \centering
    \begin{tabular}{lccc}
        \hline
        Method                            & Utilization Multiplier
                                          & Frequency Limit
                                          & Utilization at Max Speed \\
        \hline
         Static Domain Decomposition      & $1$
                                          & latency 
                                          & $\serialtime/\latency \ll 1$ \\
         Overlapping Domain Decomposition & $(1-\iterfreq\latency/\gridpoints)^\dimension$
                                          & bandwidth 
                                          & $1/(\iterfreq\latency)^\dimension \ll 1$ \\
         Domain Translation (our method)          & 1 
                                          & bandwidth
                                          & 1 \\
         \hline
    \end{tabular}
    \caption{Comparison of domain decomposition methods.}
    \label{table:method_comparison}
\vspace{-20pt}
\end{table*}
Consider each node as a serial processor holding \gridpoints{} grid points.
Each node can compute $\gridpoints-2\stencilrange$ interior grid points' next state immediately, before any network data arrives (Figure~\ref{fig:marching}).
Nodes can do this recursively for $\gridpoints/(2\stencilrange)$ steps to complete the space-time triangle with base $\gridpoints$ and height $\gridpoints/(2\stencilrange)$. This computational work can cover some or all of the latency until data arrives from a neighboring node.
As each row completes, the node records a \textit{package} of $2\stencilrange$ grid points' values that will be used by the downstream node. 

\subsection{Hardware constraints}
To quantify the performance properties of the algorithm we consider the impact of hardware components. 
Compute performance, network latency, and network bandwidth each impose an application-performance limit:

\textbf{Compute Limit}
For every package a node receives, it is able to compute the next width-$\stencilrange$ upstream facing edge of its space-time state chart. As the edge grows the diagram elongates the initial triangle into a parallelogram (Figure~\ref{fig:marching}).
Each package provides input needed for $\gridpoints$ grid-point computation steps. Assuming that it takes $\serialtime$ time to process each point update, in the compute-bound scenario the  nodes produce and consume packages at a rate of $1/(\serialtime\gridpoints)$. 

\textbf{Latency Limit}
The network can not sustain a rate greater than the equal spacing of the  $\gridpoints/(2\stencilrange)$ initially held packages in the network pipeline.
Therefore latency-bound links deliver packages at a rate of $\gridpoints/(2\stencilrange\lambda)$.
Combining the latency-limited and compute-limited expressions, shows full utilization of compute-bound performance when
\begin{equation}\gridpoints^2 > 2\stencilrange\lambda/\serialtime.\label{eq:latencylimit}\end{equation}

\textbf{Bandwidth Limit}
In addition to these constraints of compute throughput and network latency, network bandwidth bounds package transmission rate proportionally to the number of boundary points or $\gridpoints^{\dimension-1}$.  In the case of $d=1$, the bandwidth limit is  a fixed cap independent of $\gridpoints$. 

\begin{figure}[ht]
  \centering
  \includegraphics[width=0.95\linewidth]{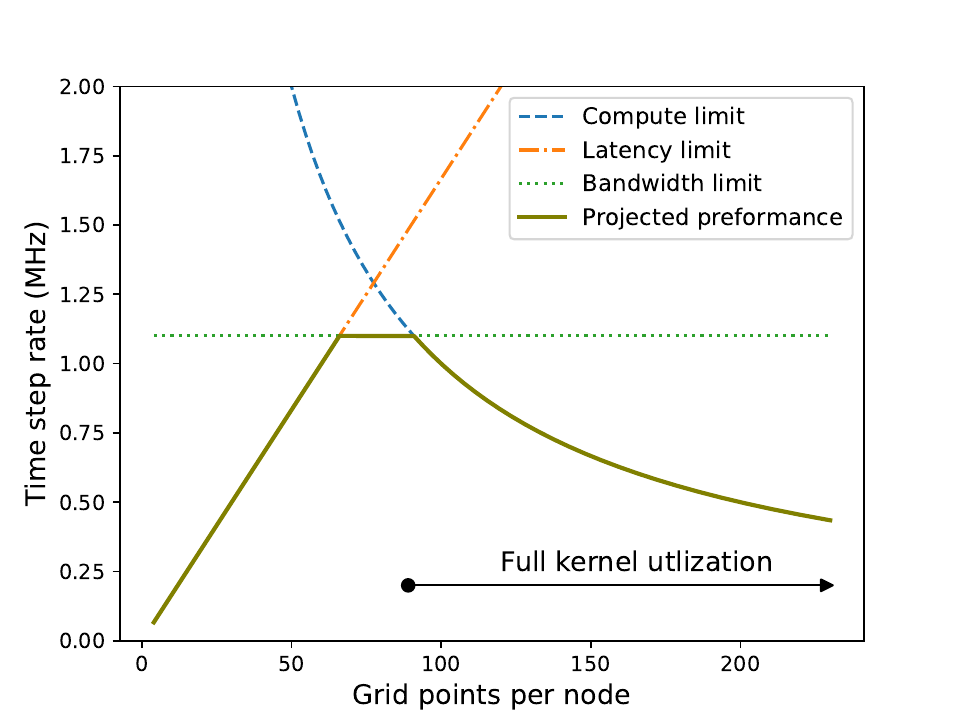}
  \caption{Time step rate can be limited by the network latency, bandwidth, or compute throughput. If the number of grid points per node is sufficiently large, the network effects are fully hidden and the kernels are expected to run at full computational utilization.}
  \Description{Time step rate vs. latency, bw, compute}
  \label{fig:utilization}
\vspace{-15pt}
\end{figure}

Figure~\ref{fig:utilization} illustrates the combined effect of network and computational constraints.  If the number of grid points per node is small, the simulation is network latency dominated.  As we increase this number, the constraint shifts to either bandwidth, or computational throughput.  In the latter case, both bandwidth and latency of the network are fully hidden, allowing the nodes to run at full kernel efficiency. 

\subsection{Dataflow Approach}

There is inherent symmetry in the way that this method mimics the principle of locality in physics on computer hardware.  Spatial architectures differ from Von Neumann architectures in the topology of the memory/processor system. 
In Von Neumann architectures, there is a memory hierarchy with transmission between processors typically done by shared memory access. The wire density significantly limits access rates to main memory.  Further, being tied to the same memory pool at all times with all processors limits the asynchronous capabilities of the system.  That is, processors have to be working on the same time step (or agreed chunk of time) within the same execution pathway.  

Spatial architectures offer a compelling alternative to traditional von Neumann-based processors for scientific workloads that exhibit high degrees of data parallelism and spatial locality. 
In spatial architectures main memory is distributed evenly with the processors and is at the nano scale. This allows memory bandwidth to be perfectly matched to processor speed.  Processors are directly connected with a Network-on-Chip (NoC) router (also at nano scale) which allows data to be transmitted in parallel with low energy penalty in a very rapid, but reliable transmission. Thus the processor/memory topology is completely flat without memory hierarchy.
Computation is explicitly mapped onto an array of processing elements (PEs). The distributed memory model removes synchronization bottlenecks related to memory accesses, which frequently limit performance in conventional HPC systems. By colocating data and computation, spatial architectures enable efficient use of memory bandwidth and energy, particularly for compute and memory intensive kernels commonly found in numerical simulation, finite-element methods, and stencil computations.


The NoC connectivity allows for asynchronous, decentralized execution. Each PE operates independently, performing computations and exchanging data with a fixed set of neighbors without global synchronization. This neighbor-to-neighbor communication model supports fine-grained parallelism and allows data to stream through the compute fabric while intermediate results are processed in-place. As a result, spatial architectures can sustain high throughput and deterministic execution patterns—attributes critical for large-scale simulations and time-stepped solvers. Their ability to exploit both spatial and temporal locality makes them well-suited for accelerating structured grid computations and other spatially-amenable problems in scientific computing. As we will show in the next section, these features of spatial architectures allow an elegant implementation of the Domain Translation algorithm.


To implement Domain Translation, we can take advantage of the spatial hardware architecture by tilting the calculation plane in space-time by 45 degrees (see Figure~\ref{fig:marching}). This is similar to maintaining the concept of ``now'' on a Penrose diagram as light moves through space time.  On hardware, data is shifted up and to the left in time while the space relative to ``now'' is always left/right.  This ensures that all data movements are at most 2 processor hops away in the case of time evolution and 1 hop in space.  In this way, all relevant data lives in the local neighborhood of the processor operating on it at all times, regardless of when ``now'' is.  Said another way, the data and instruction execution pathway are always coincident and local on all processors at every execution cycle at every point in the simulation (regardless of how many wafers are involved).  Further, there is no need for host-device interaction; the entire cluster execution is self orchestrated after compile time.  Finally, applying similar concepts to relativity in computing allows us to segment ``now'' across multiple wafers at once such that our time horizon extends past the latency barrier between wafers.  As long as the calculation time horizon on each wafer is larger than the network latency, the latency can be completely hidden and allow efficient scaling to large clusters. 

\section{Implementation on Wafer Scale Engine}
This section describes the domain translation software framework and the hardware used in our experiments. 

\subsection{Stencil Implementation}
We implemented a generic stencil-code framework in the Tungsten dataflow language \cite{santos2024breaking}.
The framework allows declaring state variables,
defining time-stepping operators,
and using translation primitives to move data by a $(\stencilrange,\stencilrange)$-step in the grid space with a combination of memory copy and network transmission.

The framework itself is lean and consists of 1,000 lines of code, including core-to-core and node-to-node communication and all kernel functions.

The code is parametrized by the number grid points ($n\times n$), the compute grid dimensions $(\fabriclen_\mathrm{w},\fabriclen_\mathrm{h})$, and the amount of node-to-node interconnection bandwidth ($q$) to allocate. We specify these parameters at compile time.
To load the program on a cluster, we provide a graph of nodes (identified by their network addresses) for establishing software-defined network links that connect adjacent subdomains.
The graph has annotations with initial condition data for each subdomain.

The framework initiates program execution by setting all nodes to run. This is an inherently asynchronous process. For consistent performance measurements, we synchronize the wafers and cores before beginning the time stepping loop.

\begin{table}[H]
\centering
\begin{lstlisting}[style=customc]
sp socket right,up,down,left; // Communication sockets

// Grid size n x n
// p = stencil width = translation distance
xp param n,p;
xp const w = 2*p;  // Width of communication layer
sp x[n+w][n+w];    // Solution
sp y[n+w][n+w];    // Temporary to hold next timestep

// Stencil weights:  5-point Laplacian in space,
// forward Euler integration in time
sp avec[5]; // a(-1,0), a(0,-1), a(1,0), a(0,1), a(0,0)         

function sendrecv() {
  parallel {
    // Send right and up
    ∀i ∈ [w,n+w) ∀j ∈ [n,n+w) right[] <- x[i][j];
    ∀i ∈ [n,n+w) ∀j ∈ [w,n+w) up[]    <- x[i][j];
    // Receive from left and from below
    ∀i ∈ [w,n+w) ∀j ∈ [0,w)   x[i][j] <- left[];
    ∀i ∈ [0,w)   ∀j ∈ [w,n+w) x[i][j] <- down[];
  }
  parallel {
    // Send corner data received from left to above
    ∀i ∈ [n,n+w) ∀j ∈ [0,w) up[]    <- x[i][j];
    // Receive corner data from below
    ∀i ∈ [0,w)   ∀j ∈ [0,w) x[i][j] <- down[];
  }
}

function compute() {
  let i ∈ [0,n);
  let j ∈ [0,n);
  ∀i ∀j y[i+w][j+w] <- avec[0]*x[i+p  ][j+p-1];
  ∀i ∀j y[i+w][j+w] <+- avec[1]*x[i+p-1][j+p  ];
  ∀i ∀j y[i+w][j+w] <+- avec[2]*x[i+p+1][j+p  ];
  ∀i ∀j y[i+w][j+w] <+- avec[3]*x[i+p  ][j+p+1];
  ∀i ∀j y[i+w][j+w] <+- avec[4]*x[i+p  ][j+p  ];
  ∀i ∀j x[i+w][j+w] <- y[i+w][j+w];
}

function innerloop(sp niter) {
  sp iter = 0.0;
  
  while(iter < niter) {
    sendrecv();
    compute();
    iter <+- 1.0;
  }
}
\end{lstlisting}
\caption{Main time-stepping loop for the 5-point stencil heat equation code, including complete communication and computation functions. The \emph{parallel} clause expresses all statements in the block may in parallel. This allows concurrent communication in different directions and is supported by hardware micro-threads. The $\forall$-statements is a compact loop notation, which the compiler uses to infer generation of vector-instructions.}
\label{tab:he5-code}
\vspace{-10pt}
\end{table}

We implemented the Heat Equation using 5-point and 9-point central difference schemes (50 lines of code), and the Shallow Water Equations (700 lines of code) to characterize the performance of the Domain Translation algorithm on the WSE cluster. The listing for the core functions in the 5-point heat equation code is shown in Table~\ref{tab:he5-code}. The different equations differ in their variable counts and arithmetic intensities. By varying the types of loads we are better able to characterize performance.

\subsection{Numerical Methods}
\begin{table}
    \centering
    \begin{tabular}{lrr}
        \hline
        System & Field Vars & FLOPs \\
        \hline
        HE, 5-point stencil   &   1 &   9 \\
        HE, 9-point stencil   &   1 &  17 \\
        SWE (even half-step)  &   3 &  94 \\
        SWE (odd  half-step)  &   4 &  61 \\
        SWE (full time-step)  &   7 & 155 \\
        \hline
    \end{tabular}
    \caption{Hyperparameters of analyzed workloads}
    \label{tab:stencilhyper}
\vspace{-10pt}
\end{table}

\subsubsection{Heat Equation}
Five-point stencils can numerically solve second order PDEs of the form $\dot u = A u_{xx} + C u_{yy} + D u_x + E u_y$.
A linear combination of each point's neighbors provide local approximations for the directional derivatives.
Adding one to the central coefficient and scaling all terms by a finite $\Delta{t}$ implements Eulerian time integration
via recursively replacing its input.
We implemented and thoroughly characterized a generic five-point stencil with fixed coefficients.
The heat equation is a familiar example that fits this form.
Each iteration therefore performs five multiplications and four additions for a total nine FLOPs per point per step (Table~\ref{tab:stencilhyper}). The 9-point stencil is similar, except it uses 8 immediate neighbors of each grid point in the computation, and performs 17 FLOPs per grid point per time-step.

\subsubsection{Shallow Water Equations}

The shallow water equations (SWE) define a system of non-linear hyperbolic conservation laws. These are PDEs that model inviscid fluid flow for systems where length scale is much greater than the depth scale.
They result from depth integration of the incompressible Naiver-Stokes equations.
SWE models velocity and fluid height subject to gravity, Coriolis force, and oceanic topography. 
Our simulation uses the Mercator projection so that the grid has a flat topology. In the resulting latitude-longitude coordinates, SWE takes the form
\begin{align*}
&\frac{\partial u}{\partial t} + \mathbf{v}\cdot\nabla u - \left( \coriolis + \frac{u}{a} \tan\latitude \right) v + \frac{g}{a\cos\latitude}\frac{\partial h}{\partial \longitude}=0,\\
&\frac{\partial v}{\partial t} + \mathbf{v}\cdot\nabla v + \left( \coriolis + \frac{u}{a} \tan\latitude \right) u + \frac{g}{a}\frac{\partial h}{\partial \latitude}=0,\\
&\frac{\partial s}{\partial t} + \mathbf{v}\cdot\nabla s + \frac{s}{a\cos\latitude}\left( \frac{\partial u}{\partial \longitude} + \frac{\partial[v\cos\latitude]}{\partial \latitude} \right)=0,
\end{align*}
where $g$ is the gravitational acceleration, $a$ is the Earth's radius, $\coriolis$ is the Coriolis coefficient, $\longitude$ is the longitude, $\latitude$ is the latitude, and $\mathbf{v} = (u, v)$.

The SWE field variables are surface velocity $(u,v)$ and water surface height $h$.
The method uses a constant field for oceanic topography $b$ defined as the distance from Earth's center and represented relative to a reference sphere with radius just below the lowest abysmal point to avoid loss of precision. The variable $s$ denotes the water depth, $s=h-b$.
Additionally, we maintain constant fields: a shoreline indicator field encodes the water boundary to enforce a no-slip condition;
pre-computed sine, cosine, and secants (of latitude) allow the method to work in latitude-longitude coordinates.

We use a Lax-Wendroff spatial discretization and a two-stage Runge-Kutta (RK2) time integration scheme.
This uses \textit{cell center} half steps in space and time to achieve a second-order estimate of field evolution.
The second-order method provides conservation, higher accuracy over longer intervals and stability for hyperbolic PDEs. This permits taking larger time steps. Indeed, solving hyperbolic problems requires a time integrator that includes an interval of the imaginary axis as part of its stability region in the plane of complex numbers. This disqualifies Eulerian time-stepping approaches.

The first half step has 94 FLOPs and computes 4 neighborhood fields.
The second half step has 61 FLOPs and computes 3 neighborhood fields.
In total, each full time-step has 155 FLOPs, computing 7 neighborhood fields, and remaps the grid twice.

\subsection{The Wafer-Scale Engine Cluster} \label{sysoverview}
A wafer-scale engine (WSE) \cite{WSE_HPC_2} is a system built with the largest processor chip ever produced, and is an example of a spatial architecture.
A WSE comprises a 2D grid of tiles, each composed of a processing element (PE), local memory, and a router.
Each router has five ports, connecting one each to its immediate cardinal neighbors with less than  2 ns latency, and one to the local PE. Routers forward 32-bit messages called \textit{wavelets} on 24 virtual channels.
A virtual channel multicasts traffic arriving from any one of the five ports to any subset of the five ports. Routers can be dynamically reconfigured in response to specially formatted control wavelets.

Specialized tiles designated for off-wafer IO live along the left and right edges of the wafer.
In total, there are 132 IO tiles, with 66 per side on the left and right.



\begin{figure}[h]
  \centering
  \includegraphics[clip, trim=8cm 4cm 8cm 3cm, width=1.0\linewidth]{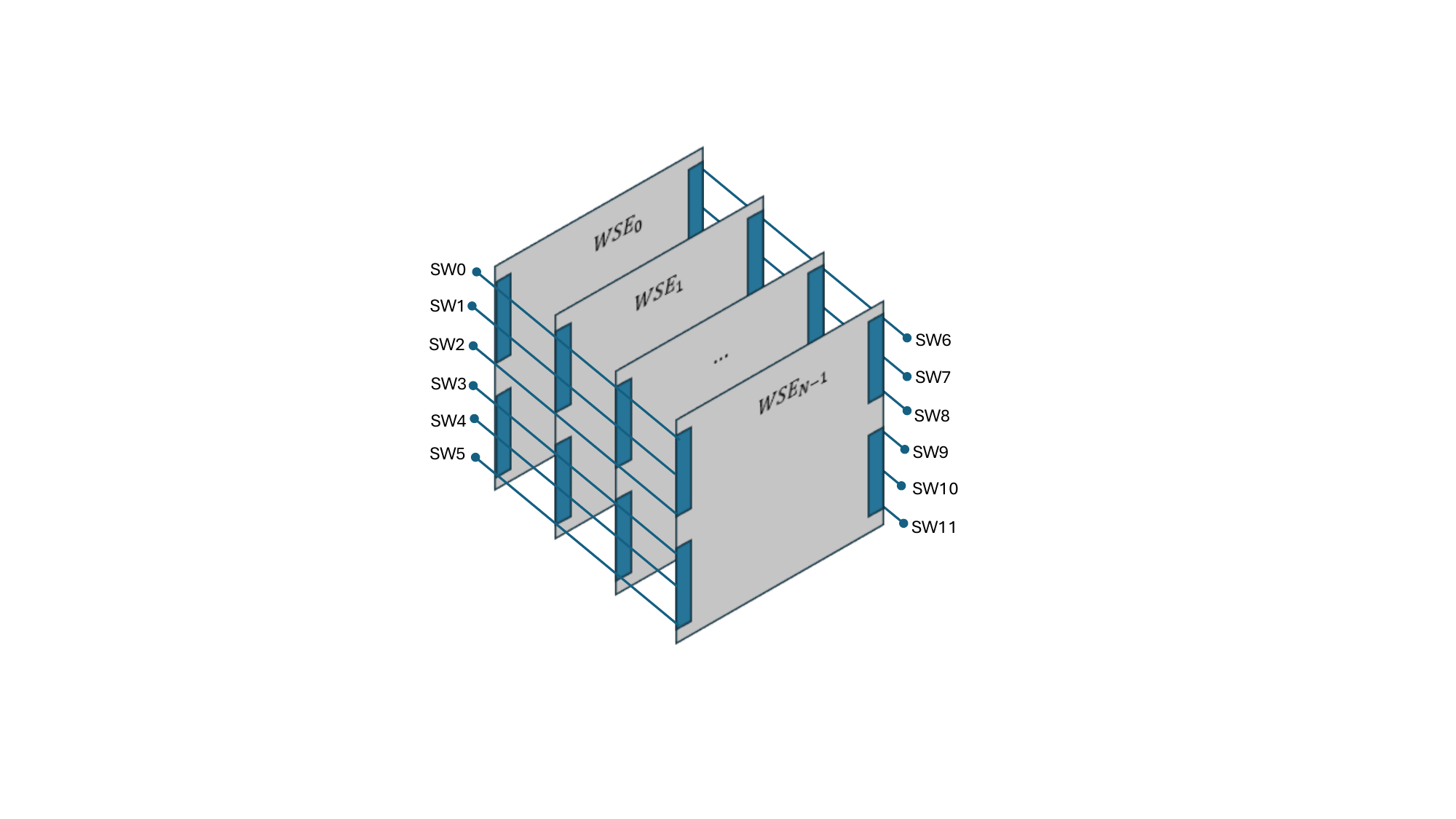}
  \caption{Switch Topology for an arbitrary number of WSEs. The same port index for all wafers in a cluster live on the same network switch. }
  \Description{High-level View of WSE}
  \label{fig:nettopology}
\end{figure}

We used a cluster of 64 WSE nodes for our experiments.
In the cluster, the $i^\mathrm{th}$ wafer's $j^\mathrm{th}$ port connects to the $j^\mathrm{th}$ switch's $i^\mathrm{th}$ port (Figure~\ref{fig:nettopology}).
Each node has twelve 100 Gbps ports, thus with a collection of twelve 64-port switches, the wafers are fully interconnected.

\begin{figure}[ht]
  \centering
  \includegraphics[width=0.6\linewidth]{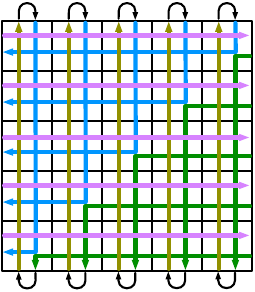}
  \caption{On-wafer routing. Horizontal flow (purple) enters the node's left edge and creates a daisy chain between every pair of horizontally adjacent cores. It egresses the system’s right edge. Vertical flow (dark green) enters from the right edge and makes a turn toward the bottom edge. The bottom row of cores receive this message and retransmit it upwards in a daisy chain (light green) among every pair of vertically adjacent cores. The top row of cores transmit the message back down (blue) with a turn to egress the node at its left edge. This scheme uses 8 virtual channels. Both daisy chains use two virtual channels alternating channels on an even/odd scheme. Both turns use two virtual channels with the change occurring at the bend.}
  \Description{On-wafer routing.}
  \label{fig:routing}
\end{figure}

Wafer communication ports are on the right and left fabric edges.
The program's core array receives data from its left and bottom edges and transmits toward the top and right.
Accordingly, the horizontally directed traffic receives from the wafer's left edge and sends to the right.
The vertically directed traffic routes with a 90-degree turn send to the left edge and receive from the right edge (Figure~\ref{fig:routing}).

At the largest compute grid size $w_h \sim w_w \sim 1,000$, the node-to-node transmission involves up to 1,000 fabric hops on the wafer to get to the edge, an ethernet link through a switch to the next wafer, and up to 1,000 fabric hops to get to the receiving tile. The ethernet routing latency is close to $10 \mathrm{\mu{s}}$ and 
the average link latency for using the sequential node performance model (Section~\ref{sec:perf_assessment}) is about $1000^{-1}(10 \mathrm{\mu{s}} + 2000 \mathrm{ns}) = 12 \mathrm{ns}$. The \algorithm algorithm allows us to benefit from the low average link latency whereas ghost methods are limited by the worst-case $10 \mathrm{\mu{s}}$ latency.

\begin{figure}[ht]
  \centering
  \includegraphics[width=0.75\linewidth]{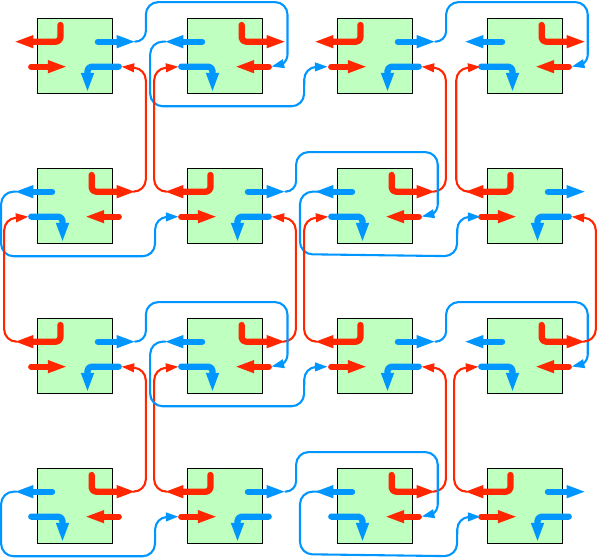}
  \caption{An example cluster topology showing the original spatial stencil and its Y-axis mirror which alternate between every wafer in the cluster. Mirroring enables data to enter and exit through the same switch, minimizing communication latency.}
  \Description{Cluster connectivity.}
  \label{fig:cluster}
\end{figure}

The physical wafer interconnection does not match the desired application interconnection: physically, every wafer's right-hand side is connected, whereas the application needs to send from the rightmost extent of its grid to the leftmost extent of an adjacent node's grid.
To handle this we compile the program as two mirror-image variants.
The nodes are assigned variants in a checkerboard pattern (Figure~\ref{fig:cluster}).
The checkerboard pattern constrains cluster to having multiples of four nodes.

\section{Performance Characterization}
This section describes our performance modeling and the specific experiments we ran. As we will show, the models closely match the experiments at all scales.

\subsection{Methodology}
\label{sec:perf_assessment}
To assess scaling and fit a performance model, we ran a sweep of performance tests varying weak scale, strong scale, and node performance (Table~\ref{tab:perfsweep}). At regular timestep intervals, all cores record their 48-bit hardware clock-cycle counter to local memory.
Instrumented runs produce an array with dimensions $(\mathrm{cluster_w} \times \mathrm{cluster_h} \times \mathrm{w_w} \times \mathrm{w_h} \times \mathrm{samples})$.
For large runs (50M cores, 1000 samples) this produces $300~\mathrm{GB}$ of telemetry.
Therefore we only retain data from a subset of cores for analysis.

We fit measurements to a performance model.
The model uses the minimum of compute-bound performance and IO-bound performance. 
Because the \algorithm algorithm renders latency irrelevant, we do not include it in the model.

\begin{table*}[ht]
    \centering
    \begin{tabular}{lcll}
        \hline
        Parameter     & Dimension & Values  & Parameter Range \\
        \hline
        Weak Scale   & (nodes)       & $1$, $2^2, 4^2, 4{\times}6, 2{\times}30, 6{\times}10, 8^2$ & 1 - 64 wafer-scale nodes \\
        Strong Scale & (points/core) & $2^2, 4^2, 8^2, 16^2, 32^2, 64^2$                          & 4 - 4k grid points per core \\
        Node Size    & (core/node)   & $720^2, 744{\times}1116$                                   & 500k and 900k cores per node \\
        Node Clock   & (GHz)         & $0.75, 1.2$                                                & 0.8 - 2 PFLOP/s peak performance \\
        \hline
    \end{tabular}
    \caption{ Parameters settings used in grid sweep for performance characterization. }
    \label{tab:perfsweep}
\end{table*}

We fit models to three different codes: the heat equation with 5- and 9-point stencils, and the Shallow Water equations. The compute cost of the performance models are shown in Table~\ref{tab:perfmodel}. In the limit of large $n$, all models become compute bound, and the asymptotic utilization is shown in the table.

\begin{table*}
    \centering
    \begin{tabular}{llcccc}
        \hline
        Code          & Performance model  & Flops/step & Asymptotic utilization ($n\rightarrow \infty$)& Realized utlization & I/O (words/step)\\
        \hline
        5-pt heat eq. & $f(n) = 105 + 3.74n + 6.72n^2$  & \phantom{00}$9n^2$ & 67\%  & 67\% $n=64$ & $4n+4$\\
        9-pt heat eq. & $f(n) = 97 + 3.5n + 9.37n^2$  & \phantom{0}$17n^2$ & 91\%  & 88\% $n=64$& $4n+4$\\
        SWE           & $f(n) = 1026 + 183.2n + 137.6n^2$ & $155n^2$ & 56\% & 53\% $n=24$& $64n+80$ \\
        \hline
    \end{tabular}
    \caption{Performance models for different PDE simulations. The models were fit to single node experiments not using external I/O. $f(n)$ gives the estimated cost of one timestep in clock cycles. $n$ indicate the problem size, where the number of grid points per core is $n\times n$. The model error is $<5\%$ for $n<=5$, and $<0.3\%$ for $n>5$. The utilization is measured in fraction of peak flops/s (2/cycle). The I/O columns lists how many floating points words are received each timestep.}
    \label{tab:perfmodel}
\end{table*}

We characterize a node's compute-bound performance in an IO-unconstrained environment.
Here, on-wafer routing directly bridges opposite edges of the core array.
We regress iteration time's dependence on grid points per core as $\itertime = \sum_{i=0}^{\dimension} a_i\gridpoints^i$. The resulting models are presented in Table~\ref{tab:perfmodel} and are accurate to a fraction of a percent for $n>5$.

IO-bound performance is assessed by measuring link payload per iteration.
We independently regress horizontal and vertical payloads as $(b_0, b_1, b_2)\cdot(\gridpoints\fabriclen_\mathrm{w}, \fabriclen_\mathrm{w}, 1)$ and $(d_0, d_1, d_2)\cdot(\gridpoints\fabriclen_\mathrm{h}, \fabriclen_\mathrm{h}, 1)$.
The I/O model assumes that the left and right side of the wafer each has 300 Gbit/s network bandwidth in each direction. Table~\ref{tab:perfmodel} lists the amount of I/O between a core and its neighbors for each code.

The current implementation has two IO limitations:
\begin{enumerate}
    \item
    In their vertical transmissions each core sends its horizontal neighbor points.
    At the node-to-node level this is redundant.
    A periodic data filter on node transmit coupled with data replication upon receive would achieve $b_1=0$.
    This is significant for our implementation because we have large $\fabriclen$ and small $\gridpoints$.
    \item 
    We provision dedicated vertical and horizontal communication channels each with the same total number of IO links.
    This means that the maximum of horizontal and vertical payloads limits performance.
    With optimization the average (not the max) will limit performance.
\end{enumerate}

In the strong scaling limit (small $n$) these limitations can have significant impact, and resolving both of them could improve timstepping rate by up to 50\% for $n=2$.

\subsection{Measured results}
\subsubsection{Characterization of 5-point heat equation}
\begin{figure}[ht]
  \centering
  \includegraphics[width=0.9\linewidth]{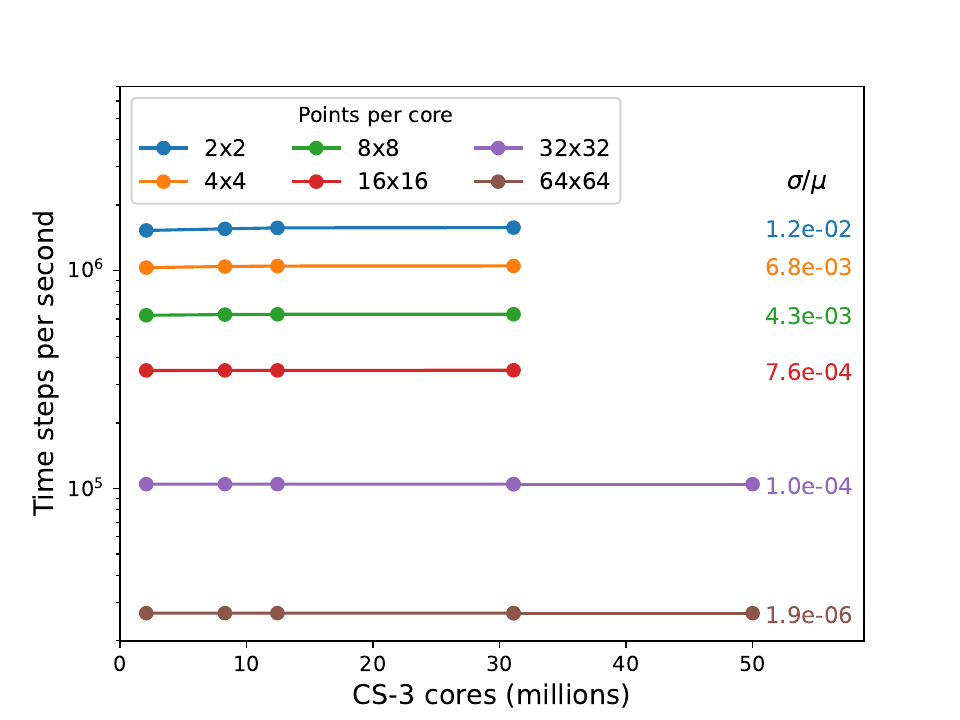}
  \caption{Weak scaling of the heat equation for runs on 4 to 60 CS-3 node counts. The observed performance was remarkably consistent for each points/core setting.  }
  \Description{Scaling.}
  \label{fig:weak_scaling}
\end{figure}

The 5 point heat equation is the most data-intensive. For that reason, we choose it for our main measurement and characterization. We conducted a range of experiments on a cluster of wafer-scale CS-3 systems~\cite{CS3} running at a 750MHz clock frequency.  We varied the number of nodes and the number of grid points per core (Table~\ref{tab:perfsweep}).  We used the resulting measurements to assess the weak scaling of the performance measured in time steps per second.  While varying the number of CS-3 nodes from 4 to 64, we observed near-perfect weak scaling with a remarkably consistent performance for each setting of grid point number per core (Figure~\ref{fig:weak_scaling}).

The observed performance was very consistent between runs with extremely low relative variances of measurements with different node counts. We observed weak scaling efficiencies ranging from 98.8\% ($\gridpoints=2$) to 99.9998\% ($\gridpoints=64$) when scaling from 4 to 60 nodes. It is reasonable to ask whether the experimental setup has the fidelity to report a result to six significant figures. The measurement is a hardware clock cycle counter (accurate to the nanosecond) for a period of 255,000 time steps with each contributing over 1,000 cycles. Therefore, the raw data has fidelity to 9 figures.

\begin{figure}[ht]
  \centering
  \includegraphics[width=0.9\linewidth]{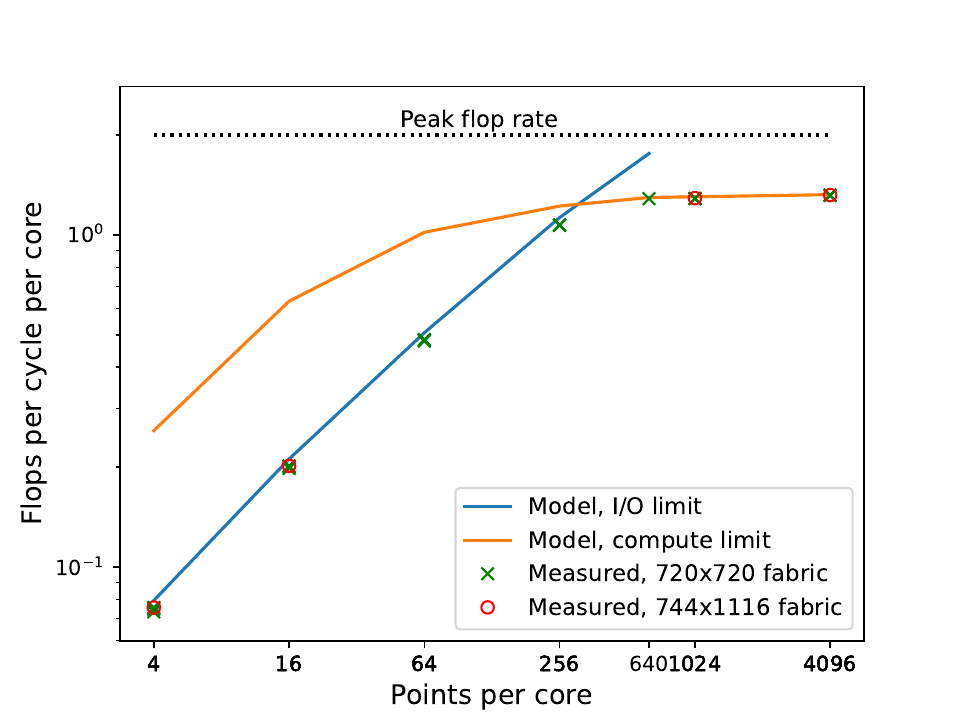}
  \caption{Comparison of model prediction with measured results.  The model correctly shows the cross-over between IO-bound and compute-bound behavior.}
  \Description{Model.}
  \label{fig:model}
\end{figure}

For all experiments that we conducted, the number of grid points on each CS-3 system was sufficient to fully hide the cross-node communication latency, but the specific choice resulted in either compute-bound or IO bandwidth-bound behavior.   We applied the model described in Section~\ref{sec:perf_assessment} and compared model predictions to the measured data.  Figure~\ref{fig:model} shows this comparison expressed as FLOP/core/s.  We can see that using fewer than 256 points per core results in communication-bound behavior.  As we use more points per core, the computation cost grows faster than that of the communication and the computation reaches its compute bound.  In the compute-bound regime we observe up to 1.32 flops per cycle, which is 66\% of the single precision peak of 2 flops per cycle on the CS-3 system. 

Having established a strong alignment of model predictions with measured data, we can plot strong scaling. The top panel of Figure~\ref{fig:strong_scaling} shows our performance model, including I/O, for scaling problems of various sizes across 4 to 128 CS-3 nodes, as well as data points measured on cluster simulations. We see close agreement between our model and real measured performance.

\subsubsection{9-point heat equation and peak performance}
\begin{figure}[ht]
  \centering
  \includegraphics[width=0.83\linewidth]{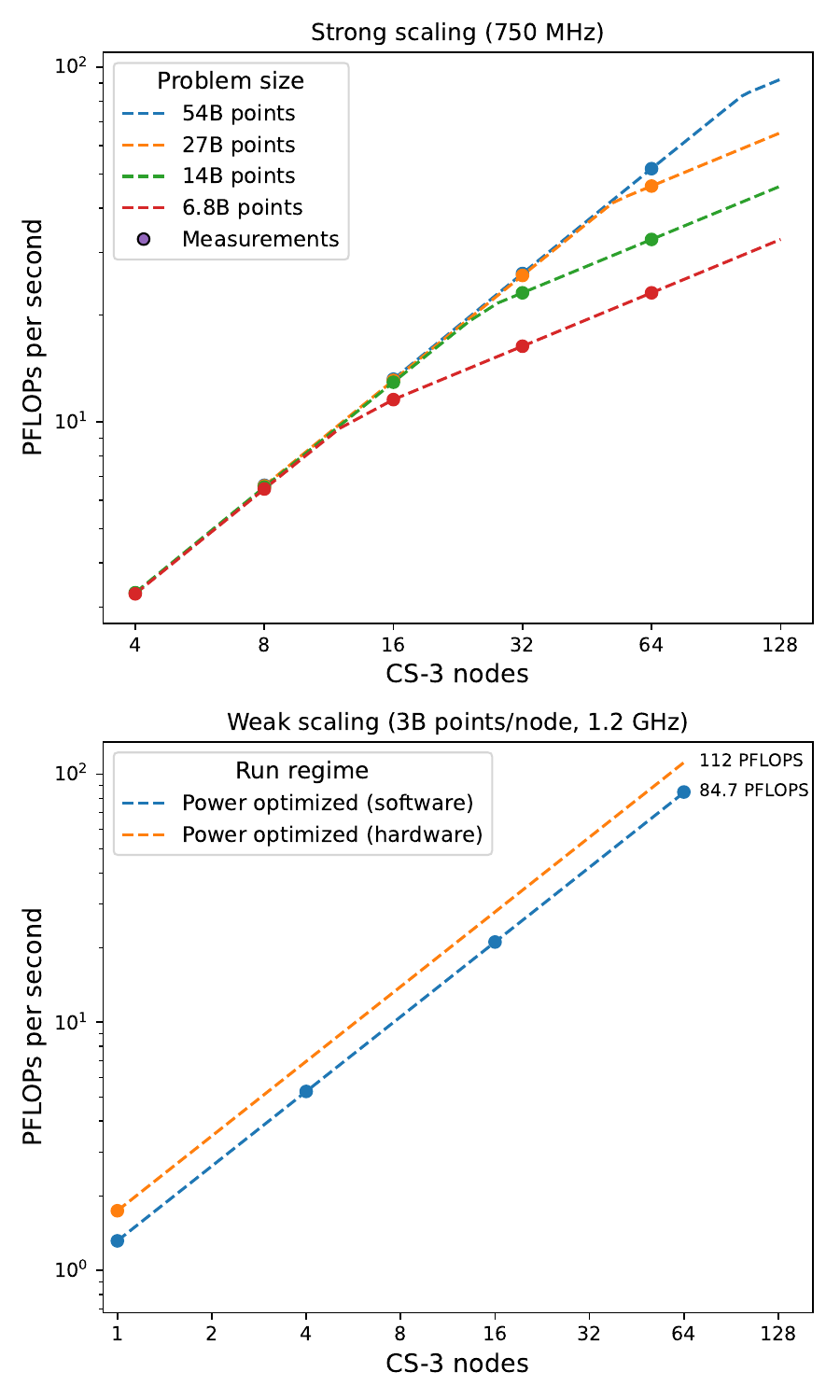}
  \caption{Top: strong scaling for the 5-point heat equation code on different number of WSE nodes, configured at a 750 MHz clock frequency. It shows perfect scaling up to 64 nodes for the largest simulations, and very close agreement with the performance model. Bottom: weak scaling for 9-point heat equation on system at 1.2GHz. \emph{Blue line}: code optimized to reduce power throttling at the higher clock frequency. On 64 nodes and 192 billion grid points, we achieved 84.7 PFlops. \emph{Yellow line}: system with enhanced power supply. }
  \Description{Strong scaling}
  \label{fig:strong_scaling}
\vspace{-15pt}
\end{figure}

Our systems allow configuring the clock frequency. We ran experiments at 750MHz and 1.2GHz. At the 1.2GHz operating point, the system is power constrained and can throttle significantly. We created a power optimized version of the 9-point heat equation that takes advantage of a small region of power efficient memory in each core. The bottom panel of Figure~\ref{fig:strong_scaling} weak scaling performance for this code operating at 1.2GHz with 3 billion grid points per node. The weak scaling behavior is nearly perfect and we achieved 84.7 PFLOPS using 64 nodes. The power specification per CS-3 node is 23kW. Thus, the power-efficient performance of the 64-node CS-3 cluster measures at 57 GFLOPs per Watt. In comparison, the current leader of Green500 (JEDI) achieves 72.7 GFLOPs per Watt on a dense linear algebra workload \cite{Green500}. To the best of our knowledge, there is no sparse computation on any cluster that is near the efficiency we observe with the CS-3 cluster.

At the 1.2GHz operating point, even the power optimized code is affected by power throttling. To further push the performance, we ran the 9-point heat equation code on a special node outfitted with an enhanced power supply. This allowed continuous unconstrained operation at 1.2GHz without power throttling. In this setting the code performed over 2.1 GFLOPS per core, and at 88\% of peak performance. At 64 nodes of scale, this performance would produce 112 PFLOPS given our perfect weak scaling. 


\subsubsection{The Shallow Water Equations}
The heat equation is a simplistic model. To explore the efficacy of the Domain Translation method on more complex and realistic equations, we tested the \algorithm method for the shallow water equations using the GEBCO\_2024 Grid \cite{gebco}. The data set offers a global continuous terrain model for ocean and land with a spatial resolution of 15 arc seconds or 462 meters.

\begin{figure*}[ht]
  \centering
  \includegraphics[width=0.9\linewidth]{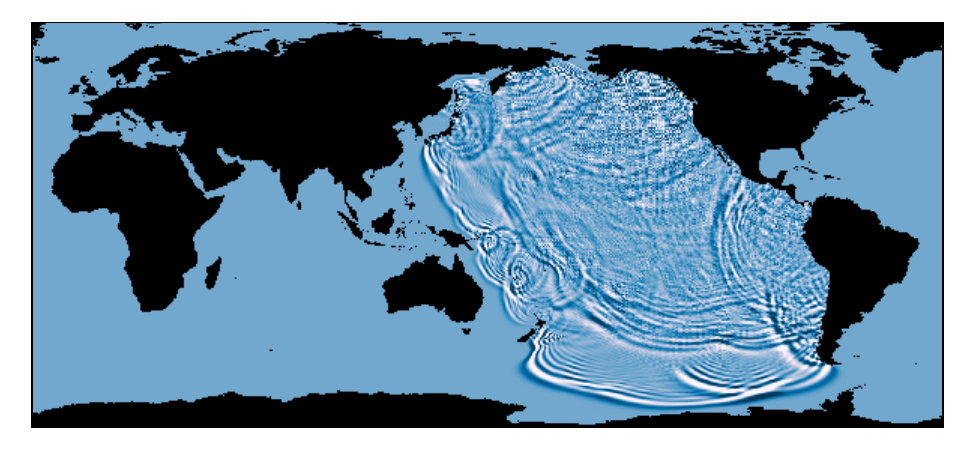}
  \caption{Planetary-scale tsunami wave propagation 14 hours after asteroid impact with energy of 2.4M tons of TNT equivalent.}
  \Description{Large-scale flow}
  \label{fig:planet}
\end{figure*}

\begin{figure}[ht]
  \centering
  \includegraphics[width=0.75\linewidth]{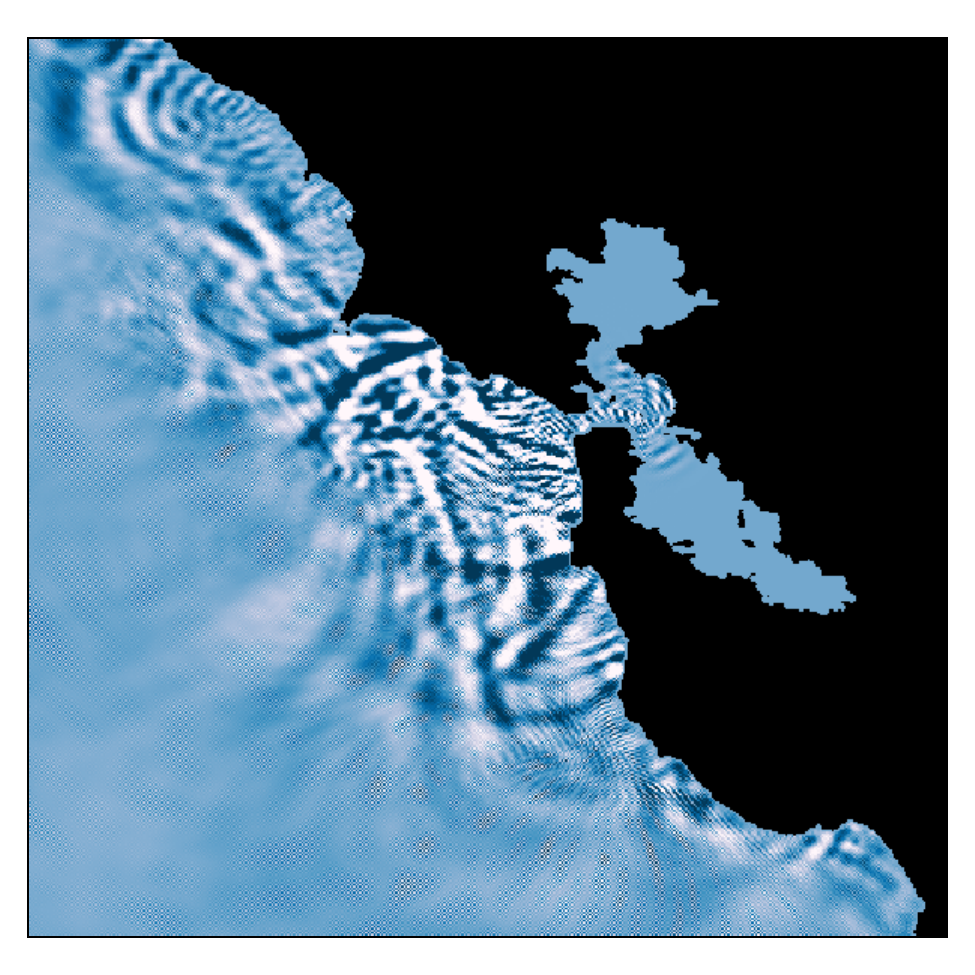}
  \caption{Wave impact at San Francisco Bay}
  \Description{Large-scale flow}
  \label{fig:SF}
\end{figure}

Figure~\ref{fig:planet} shows the outcome of asteroid impact in the ocean. The asteroid disintegrates or ends at the ocean floor. Its kinetic energy dissipates as thermal and mechanical energy in the ocean. We simulate the impact as a sine hump of water elevated above the surface of the ocean whose potential energy equals 90\% of the impact kinetic energy (2.4M tons of TNT equivalent). The sine hump is an initial condition given as input to the SWE model, which simulates the resulting waves via numerical solution of PDEs. In this example, the sine hump spans 30K km$^2$, elevates to 200 m above the ocean surface at its highest, and contains 5.8T m$^3$ of water. This simulates the energy of a 2M kg asteroid impacting at 100 km/s. Figure~\ref{fig:SF} shows the wave impact zoomed at San Francisco Bay.

The performance model for the SWE is shown in Table~\ref{tab:perfmodel}. As with the heat equation, we observed near-perfect weak scaling across multiple CS-3 nodes. In the compute bound regime we achieve 53\% of peak performance. 
We plan to extend our work to larger clusters and future generations of our hardware, and expect to observe increased performance consistent with model predictions.

\section{Conclusions}
A central implication of this work is a pathway to achieve unprecedented performance for physical system modeling.
We demonstrated that we can maintain high system efficiency on a large distributed cluster at a fine-grain parallelism.  This delivers scaling not only in the spatial domain (i.e., weak scaling) but also in the time domain (i.e., strong scaling).  This enables study of large scale systems in ways that were previously impossible or cost and time prohibitive.  This has significant implications for study of long time horizon physics, uncertainty quantification, design optimization, cyberphysical security, and real time digital twinning.  We have shown that it is possible to extend our previous research on single-device PDE solutions to multi-device with little to no deterioration in calculation speed.

The Domain Translation method can hide latencies that are quadratic in the amount of memory (given by Eq.~\ref{eq:latencylimit}). In this context, an entire cluster can be considered a unit, and its aggregate memory used to cover latency. Besides the strong scaling benefits mentioned above, a further implication of our work is that clusters in different cities could be fruitfully networked and use Domain Translation to overcome the millisecond or longer network latencies, opening the door for running parallel applications across multiple exa-scale machines.

For these reasons, we were careful to demonstrate a meaningful result for science at planetary scale. The shallow water equations solved here represent the core of several important geophysical applications.  In addition to the tsunami application described above, shallow water solvers are critical components of both atmospheric and oceanic modeling.  In these disciplines, one first constructs a shallow water model to test numerics and computational implementation, and then extends these models to three-dimensional atmosphere and ocean models used for numerical weather prediction and Earth system modeling.  For example, the state-of-the-art CESM, E3SM, FV3 and MPAS atmosphere and ocean models (\cite{Skamarock2012,Ringler2013,Small2014,Golaz_et_al:2022,Dahm2023,Donahue_et_al:2024} are derived from the shallow water models given in \cite{Putman2007, thuburn2009numerical, TayFour10}).  

The two-dimensional shallow water equations are an ideal first step in the development of these models because they capture the large scale dynamics which dominate the flow and the wave propagation behavior that controls the timestep.  To develop a full atmosphere model from a shallow water code, one first adds vertical layers referred to as {\em stacked shallow water}, followed by vertical coupling between the terms and additional prognostic variables for vertical velocity, temperature and various water species.  

Because of the 2D domain decomposition used in modern atmosphere and ocean models, the parallel performance of stacked shallow water equations is an excellent model for the performance of the full equation set.  This is because the nearest neighbor communication patterns  will be nearly identical and the ratio of floating point operations to communication will be quite similar.   Evolving a stacked shallow water model into a full atmospheric model involves many additional prognostic variables and associated computations, but the extra work is all on-processor, resulting in a slower model but with improved strong scaling.  The stacked shallow water system represents the core computational bottleneck in modern numerical weather prediction and Earth system modeling.  

Our results running shallow water equations on the Cerebras system demonstrate the promise of this architecture for atmosphere and ocean models critical for weather prediction and Earth systems assessments.
This sets the stage for greatly enhanced Earth system modeling with an order of magnitude increased throughput and 1.5 order of magnitude improved power efficiency.

\begin{acks}

The authors would like to thank Amirali Sharifian and Milad Hakimi for their work on improving the compiler to support this project. 

Sandia National Laboratories is a multimission laboratory managed and operated by National Technology and Engineering Solutions of Sandia LLC, a wholly owned subsidiary of Honeywell International Inc. for the U.S. Department of Energy’s National Nuclear Security Administration under contract DE-NA0003525.

This work was supported in part by the U.S. Department of Energy, Office of Science, Office of Advanced Scientific Computing Research's Computer Science Portfolios program.

This work was also funded by the United States Department of Energy, National Energy Technology Laboratory. Neither the United States Government nor any agency thereof, nor any of their employees, makes any warranty, express or implied, or assumes any legal liability or responsibility for the accuracy, completeness, or usefulness of any information, apparatus, product, or process disclosed, or represents that its use would not infringe privately owned rights. Reference herein to any specific commercial product, process, or service by trade name, trademark, manufacturer, or otherwise does not necessarily constitute or imply its endorsement, recommendation, or favoring by the United States Government or any agency thereof. The views and opinions of authors expressed herein do not necessarily state or reflect those of the United States Government or any agency thereof.

\end{acks}

\bibliographystyle{ACM-Reference-Format}
\bibliography{main}

\end{document}